\documentclass[english,twocolumn]{emulateapj}
\usepackage[T1]{fontenc}
\usepackage[latin1]{inputenc}
\usepackage{array}
\usepackage{amssymb}

\makeatletter

\providecommand{\tabularnewline}{\\}

\usepackage{array}

\makeatletter

\usepackage{times}

\newcommand{\peryr}{\mathrm{yr}^{-1}}

\shorttitle{Massive Perturber-driven interactions}
\shortauthors{Perets, Hopman and Alexander}

\makeatother

\usepackage{babel}
\makeatother
\begin{document}
\newcommand{\Ms}{M_{\star}} \newcommand{\Rs}{R_{\star}} \newcommand{\Ls}{L_{\star}}

\newcommand{\Mo}{M_{\odot}} \newcommand{\Ro}{R_{\odot}} \newcommand{\Lo}{L_{\odot}}
\newcommand{\Mbh}{M_{\bullet}} \newcommand{\np}{n_{p}} \newcommand{\Np}{N_{P}}
\newcommand{\Mp}{M_{p}} \newcommand{\Ns}{N_{\star}} \newcommand{\SgrA}{SgrA^{\star}}
\newcommand{\rMP}{r_{\mathrm{MP}}}


\title{Massive perturber-driven interactions of stars with a massive black
hole }

\author{Hagai B. Perets, Clovis Hopman\altaffilmark{1} and Tal Alexander\altaffilmark{2}}

\email{hagai.perets, clovis.hopman, tal.alexander@weizmann.ac.il}

\affil{Faculty of Physics, Weizmann Institute of Science, POB 26, Rehovot
76100, Israel}

\altaffiltext{1}{also Leiden Observatory, P.O. box 9513, NL-2300
RA Leiden } \altaffiltext{2}{The William Z. \& Eda Bess Novick
career development chair}

\begin{abstract}
We study the role of massive perturbers (MPs) in deflecting stars
and binaries to almost radial ({}``loss-cone'') orbits, where they
pass near the central massive black hole (MBH), interact with it at
periapse, and are ultimately destroyed. MPs dominate dynamical relaxation
when the ratio of the 2nd moments of the MP and star mass distributions,
$\mu_{2}\!\equiv\!\left.N_{p}\left\langle M_{p}^{2}\right\rangle \right/N_{\star}\left\langle M_{\star}^{2}\right\rangle $,
satisfies $\mu_{2}\!\gg\!1$. We compile the MP mass function from
published observations, and show that MPs in the nucleus of the Galaxy
(mainly giant molecular clouds), and plausibly in late type galaxies
generally, have $10^{2}\!\lesssim\!\mu_{2}\!\lesssim\!10^{8}$. MPs
thus shorten the relaxation timescale by $10^{2-7}$ relative to 2-body
relaxation by stars alone. We show this increases by $10^{1-3}$ the
rate of \emph{large}-periapse interactions with the MBH, where loss-cone
refilling by stellar 2-body relaxation is inefficient. We extend the
Fokker-Planck loss-cone formalism to approximately account for relaxation
by rare encounters with MPs. We show that binary stars--MBH exchanges
driven by MPs can explain the origin of the young main sequence B
stars that are observed very near the Galactic MBH, and increase by
orders of magnitude the ejection rate of hyper-velocity stars. In
contrast, the rate of \emph{small}-periapse interactions of single
stars with the MBH, such as tidal disruption, is only increased by
a factor of a few. We suggest that MP-driven relaxation plays an important
role in the 3-body exchange capture of single stars on very tight
orbits around the MBH. These captured stars may later be disrupted
by the MBH via tidal orbital decay or direct scattering into the loss
cone; captured compact objects may inspiral into the MBH by the emission
of gravitational waves from zero-eccentricity orbits. 
\end{abstract}


\section{Introduction}

\label{s:intro}

There is compelling evidence that massive black holes (MBHs) lie in
the centers of all galaxies \citep{fer+00,geb+03,shi+03}, including
in the center of our Galaxy \citep{eis+05,ghe+05}. The MBH affects
the dynamics and evolution of the galaxy's center as a whole (e.g.
\citealt{bah+76}) and it also strongly affects individual stars or
binaries that approach it. Such close encounters, which may be extremely
energetic, or involve non-gravitational interactions, or post-Newtonian
effects, have been the focus of many studies (see review by \citealt{ale05}).
These processes include the destruction of stars by the MBH, either
by falling whole through the event horizon, or by being first tidally
disrupted and then accreted (e.g. \citealt{ree88}); tidal scattering
of stars on the MBH \citep{ale+01b}; the capture and gradual inspiral
of stars into the MBH, accompanied by the emission of gravitational
waves or by tidal heating (e.g. \citealt{ale+03b,ale+03a}); or dynamical
exchange interactions in which incoming stars or binaries energetically
eject a star tightly bound to the MBH and are captured in its place
very near the MBH (e.g. \citealt{ale+04,gou+03}).

The interest in such processes is driven by their possible implications
for the growth of MBHs, for the orbital decay of a MBH binary, for
the detection of MBHs, for gravitational wave (GW) astronomy, as well
as by observations of unusual stellar phenomena in our Galaxy, e.g.
the puzzling young population of B-star very near the Galactic MBH
\citep{eis+05}, or the hyper-velocity B stars at the edge of the
Galaxy \citep{bro+05,fue+06,bro+06a}, possibly ejected by 3-body interactions
of a binaries with the MBH \citep{hil91}.

Here we focus on close encounters with the MBH whose ultimate outcome
({}``event'') is the elimination of the incoming object from the
system, whether on the short infall (dynamical) time, if the event
is prompt (e.g. tidal disruption or 3-body exchange between a binary
and the MBH), or on the longer inspiral time, if the event progresses
via orbital decay (e.g. GW emission or tidal capture and heating).
Such processes are effective only when the incoming object follows
an almost zero angular momentum ({}``loss-cone'') orbit with periapse
closer to the MBH than some small distance $q$. To reach the MBH,
or to decay to a short period orbit, both the infall and inspiral
times must be much shorter than the system's relaxation time $t_{r}$
\citep{ale+03b}. The fraction of stars initially on loss-cone orbits
is very small and they are rapidly eliminated. Subsequently, the close
encounter event rate is set by the dynamical processes that refill
the loss-cone.

The loss-cone formalism used for estimating the event rate (\citealt{fra+76,lig+77,coh+78})
usually assumes that the system is isolated and that the refilling
process is 2-body relaxation. This typically leads to a low event
rate, set by the long 2-body relaxation time.

Two-body relaxation, which is inherent to stellar systems, ensures
a minimal loss-cone refilling rate. Other, more efficient but less
general refilling mechanisms were also studied with the aim of explaining
various open questions (e.g. the stalling problem of MBH binary coalescence,
\citet{ber+05,mer+05,ber+06}, or MBH feeding, \citealt{zha+02,mir+05})
or in the hope that they may lead to significantly higher event rates
for close encounter processes. These mechanisms include chaotic orbits
in triaxial potentials \citep{nor+83,ger+85,mer+04b,hol+06} (the
presence of a MBH may however destroy the triaxiality near the center;
\citealt{mer+98,hol+02,sel02}); increased fraction of low angular
momentum orbits in non-spherical potentials \citep{mag+99,ber+06};
accelerated resonant relaxation of angular momentum near the MBH where
the orbits are Keplerian \citep{rau+96,rau+98,hop+06a,lev06}; perturbations
by a massive accretion disk or an intermediate mass black hole (IMBH)
companion \citep{pol+94,zha+02,lev+05}. Most of these mechanisms
require special circumstances to work (e.g. specific asymmetries in
the potential), or are short-lived (e.g. the IMBH will eventually
coalesce with the MBH).

Here we explore another possibility, which is more likely to apply
generally: accelerated relaxation and enhanced rates of close encounters
driven by massive perturbers (MPs). Efficient relaxation by MPs was
first suggested by \citet{spi+51,spi+53} to explain stellar velocities
in the galactic disk. MPs remain an important component in modern
models of galactic disk heating (see e.g. \citealt{vil83,vil85,lac84,jen+90,han+02}
and references therein). A similar mechanism was suggested to explain
the spatial diffusion of stars in the inner Galactic bulge \citep{kim+01}.
In addition to dynamical heating, efficient relaxation by MPs was
suggested as a mechanism for loss cone replenishment and relaxation,
both in the context of scattering of Oort cloud comets to the Sun
\citep{hil81,bai83} and the scattering of stars to a MBH in a galactic
nucleus \citep{zha+02}. \citet{zha+02} suggested MPs as a mechanism
for establishing the $M_{\bullet}/\sigma$ relation \citep{fer+00,geb+00}
by fast accretion of stars and dark matter. They also noted the possibility
of increased tidal disruption flares and accelerated MBH binary coalescence
due to MPs. In this study we investigate in detail the dynamical implications
of loss-cone refilling by MPs. We evaluate its effects on the different
modes of close interactions with the MBH, in particular 3-body exchanges,
which were not considered by \citet{zha+02}, and apply our results
to the Galactic Center (GC), where observations indicate that dynamical
relaxation is very likely dominated by MPs.

This paper is organized as follows. In \S\ref{s:outline} we present
the main concepts and procedures of our calculations. The observational
data and theoretical predictions about MPs in the inner $\sim\!100$
pc of the GC are reviewed in \S\ref{s:GC_MPs}. In section \S\ref{s:interact}
we explore the implications of relaxation by MPs for various types
of interactions with the MBH. We summarize our results in \S\ref{s:summary}.

\section{Loss-cone refilling }

\label{s:outline}

In addition to stars, galaxies contain persistent dense structures%
\footnote{Structures that persist at least as long as the local galactic dynamical
time and are substantially denser than the ambient stellar mass distribution.%
} such as molecular clouds, open clusters and globular clusters with
masses up to $10^{4}$--$10^{7}\,\Mo$. Such structures can perturb
stellar orbits around the MBH much faster than 2-body stellar relaxation
(hereafter {}``stellar relaxation''), provided they are numerous
enough. This condition can be quantified by considering a test star
randomly scattered by perturbers with masses in the interval ($M_{p},M_{p}\!+\!\mathrm{d}M_{p}$)
and number density $(\mathrm{d}N_{p}/\mathrm{d}M_{p})\mathrm{d}M_{p}$,
approaching it with relative velocity $v$ on orbits with impact parameters
in the interval ($b,b\!+\!\mathrm{d}b$).$ $ The minimal impact parameter
still consistent with a small angle deflection is $b_{\min}\!=\! GM_{p}/v^{2}$
(the capture radius), where $v$ is of the order of the local velocity
dispersion $\sigma$. Defining $B\!\equiv\! b/b_{\min}\!\ge\!1$,
the encounter rate is then \begin{eqnarray}
\left(\frac{\mathrm{d}^{2}\Gamma}{\mathrm{d}M_{p}db}\right)\mathrm{d}M_{p}\mathrm{d}b\, & \sim\left(\frac{\mathrm{d}N_{p}}{\mathrm{d}M_{p}}\right)\mathrm{d}M_{p}vb_{\min}^{2}2\pi B\mathrm{d}B\nonumber \\
 & =\frac{G^{2}}{v^{3}}\left[\left(\frac{\mathrm{d}N_{p}}{\mathrm{d}M_{p}}\right)M_{p}^{2}\right]\mathrm{d}M_{p}2\pi B\mathrm{d}B\,.\end{eqnarray}
 The total rate is obtained by integrating over all MP masses and
over all impact parameters between $b_{\min}$ and $b_{\max}$. Here
we are interested in perturbations in the specific angular momentum
$J$ of a star relative to the central MBH, and so $b_{\max}\!\sim\! r$,
the radial distance of the star from the center. MPs with substantially
larger impact parameters are much less efficient because their effect
on the MBH-star pair is tidal rather than direct.

The relaxation rate due to all MPs at all impact parameters is then

\begin{eqnarray}
t_{r}^{-1} & =\int_{b_{\min}}^{b_{\max}}\mathrm{d}b\mathrm{\int}\mathrm{d}M_{p}\left(\frac{\mathrm{d}^{2}\Gamma}{\mathrm{d}M_{p}db}\right)\nonumber \\
 & \sim\log\Lambda\frac{G^{2}}{v^{3}}\int\mathrm{d}M_{p}\left(\frac{\mathrm{d}N_{p}}{\mathrm{d}M_{p}}\right)M_{p}^{2}\,,\end{eqnarray}
 where $\log\Lambda\!=\!\log(b_{\max}/b_{\min}$) is the Coulomb logarithm
(here the dependence of $\log\Lambda$ and $v$ on $M_{p}$ is assumed
to be negligible). For stars, typically $\log$$\Lambda\!\gtrsim\!10$;
the omission of large angle scattering by encounters with $b\!<\! b_{\min}$
is thus justified because it introduces only a relatively small logarithmic
correction. This formulation of the relaxation time is equivalent
to its conventional definition (\citealt{spi87}) as the time for
a change of order unity in $v^{2}$ by diffusion in phase space due
to scattering, $t_{r}\!\sim\! v^{2}/D(v^{2})$, where $D(v^{2})$
is the diffusion coefficient.

If the stars and MPs have distinct mass scales with typical number
densities $N_{\star}$ and $N_{p}$ and rms masses $\left\langle M_{\star}^{2}\right\rangle ^{1/2}$
and $\left\langle M_{p}^{2}\right\rangle ^{1/2}$ ($\left\langle M^{2}\right\rangle \!\equiv\!\int M^{2}(\mathrm{d}N/\mathrm{d}M)\mathrm{d}M/N$),
then MPs dominate if the ratio of the 2nd moments of the MP and star
mass distributions, $\mu_{2}\!\equiv\!\left.N_{p}\left\langle M_{p}^{2}\right\rangle \right/N_{\star}\left\langle M_{\star}^{2}\right\rangle $,
satisfies $\mu_{2}\!\gg\!1$ (note that for a continuous mass spectrum,
this condition is equivalent to $-\mathrm{d}\log N/\mathrm{d}\log M\!<\!2$).

As discussed in detail in \S\ref{s:GC_MPs}, the central $\sim\!100\,\mathrm{pc}$
of the Galactic Center (GC) contain $10^{8}-10^{9}$ solar masses
in stars, and about $10^{6}-10^{8}$ solar masses in MPs such as GMCs
or open clusters of masses $10^{3}-10^{8}\,\Mo$ \citep{oka+01,fig+02,fig+04,vol+03,gus+04,bor+05}.
An order of magnitude estimate indicates that MPs in the GC can reduce
the relaxation time by several orders of magnitude, \begin{eqnarray}
\frac{t_{r,\star}}{t_{r,\mathrm{MP}}} & = & \mu_{2}\sim\frac{(N_{p}M_{p})M_{p}}{(N_{\star}M_{\star})M_{\star}}\nonumber \\
 & = & 10^{4}\left[\frac{(N_{\star}\Ms/N_{p}M_{p})}{10}\right]^{-1}\left[\frac{(M_{p}/\Ms)}{10^{5}}\right]\,.\end{eqnarray}
 Note that $\mu_{2}$ does not include possible modifications in the
value of $\log\Lambda$ for MPs due to their much larger size, which
may decrease this ratio by $O(10)$. This estimate is borne by more
detailed calculations (Fig. \ref{f:tr} and table \ref{t:models}),
using the formal definition $t_{r}\!=\! v^{2}/D(v_{||}^{2})$ with
$M_{\star}\rho_{\star}\rightarrow\int(\mathrm{d}N_{p}/\mathrm{d}M_{p})M_{p}^{2}\mathrm{d}M_{p}$
(e.g. \citealt{bin+87}, Eqs. 8-69 to 8-70). A similar result is indicated
by simulations of spatial diffusion of stars in the central 100 pc
\citep{kim+01}.

\subsection{Non-coherent loss-cone refilling}

\label{ss:noncoherent}

The Fokker-Planck approach to the loss-cone problem (e.g. \citealt{coh+78})
assumes that the effects of multiple small perturbations on the orbit
of a test star dominate over the rarer strong close encounters ($b_{\max}/b_{\min}\!\gg\!1)$,
and that the cumulative effect can be described as diffusion in phase
space. The change in the angular momentum of the test star then grows
non-coherently, $\Delta J\!\propto\!\sqrt{t}$. The change over one
orbital period $P$ is $J_{D}\!=\! J_{c}(E)\sqrt{P/t_{r}}$, where
$J_{c}\!=\!\sqrt{2(\psi-E})r$ is the maximal (circular) angular momentum
for a stellar orbit of specific relative energy $E\!=\!-v^{2}/2+\psi(r)$,
and $\psi\!\equiv\!-\phi$ is the negative of the gravitational potential,
so that $E\!>\!0$ for bound orbits. The magnitude of $J_{D}$ relative
to the $J$-magnitude of the loss-cone, \begin{equation}
J_{lc}\!\simeq\!\sqrt{2G\Mbh q}\,,\end{equation}
 determines the mode of loss-cone refilling. The relative volume of
phase space occupied by the loss-cone, $J_{lc}^{2}/J_{c}^{2}(E)$,
increases with $E$ (decreases with $r$) while $P$ decreases. Near
the MBH (high $E$) $J_{D}\!\ll\! J_{lc}$, stars diffuse slowly into
the loss-cone, and are promptly destroyed over an orbital period,
leaving the loss-cone always nearly empty. In this empty loss-cone
regime, the loss-cone is relatively large, but the refilling rate
is set by the long relaxation timescale (e.g. \citealt{lig+77}),
\begin{eqnarray}
\left(\frac{\mathrm{d}\Gamma}{\mathrm{d}E}\right)_{\mathrm{empty}} & \simeq & \frac{N_{\star}(E)}{\log(J_{c}/J_{lc})t_{r}}\nonumber \\
 & = & \frac{J_{D}^{2}(E)}{J_{c}^{2}(E)}\frac{1}{\log[J_{c}(E)/J_{lc}]}\frac{N_{\star}(E)}{P(E)}\,,\label{e:Gempty}\end{eqnarray}
 where $N_{\star}(E)$ is the stellar number density per energy interval.

Far from the MBH (low $E$) $J_{D}\!\gg\! J_{lc}$, stars diffuse
across the loss-cone many times over one orbit, and the loss cone
is always nearly full. In this full loss-cone regime the refilling
rate is set by the short orbital time, but the loss cone is relatively
small, \begin{equation}
\left(\frac{\mathrm{d}\Gamma}{\mathrm{d}E}\right)_{\mathrm{full}}\simeq\frac{J_{lc}^{2}}{J_{c}^{2}(E)}\frac{N_{\star}(E)}{P(E)}\,.\label{e:Gfull}\end{equation}
 Note that here and elsewhere we make the simplifying approximation
that the period is a function of energy only, which is true only for
motion in a Keplerian potential.

The total contribution to loss-cone refilling is dominated by stars
with energies near the critical energy $E_{c}$ (equivalently, critical
typical radius $r_{c}$) separating the two regimes (\citealt{lig+77};
see \S\ref{s:interact}). Within $r_{c}$ ($E\!>\! E_{c}$), an object,
once deflected into the loss cone, can avoid being scattered out of
it before reaching the MBH%
\footnote{In the case of inspiral, $E_{c}$ is determined by the condition $J_{D}\!=\! J_{lc}$
over the inspiral time, rather than the much shorter orbital period,
which results in a much smaller $r_{c}$ than for direct infall. Inspiraling
stars with $E\!>\! E_{c}$ can avoid being scattered directly into
the MBH before completing the orbital decay. There is no contribution
to inspiral events from regions outside $r_{c}$ ($E\!<\! E_{c}$),
since the probability of an object to remain on its low-$J$ trajectory
over the many orbital periods required to complete the inspiral, is
vanishingly small. %
}. The empty and full loss-cone regimes of infall processes can be
interpolated to give a general approximate expression for the differential
event rate for these non-coherent encounters (e.g. \citealt{you77}),\begin{equation}
\frac{\mathrm{d}\Gamma}{\mathrm{d}E}\simeq\frac{j^{2}(E)}{J_{c}^{2}(E)}\frac{N_{\star}(E)}{P(E)}\,,\label{e:dGdE}\end{equation}
 with \begin{equation}
j^{2}(E)\equiv\min\left[\frac{J_{D}^{2}(E)}{\log(J_{c}(E)/J_{lc})},J_{lc}^{2}\right]\,,\end{equation}
 where $j$ is the loss-cone limited angular momentum change per orbit,
which expresses the fact that the loss-cone can at most be completely
filled during one orbit.

\subsection{Coherent loss-cone refilling}

The loss-cone formalism can be generalized to deal with MPs in an
approximate manner with only few modifications. The capture radius
for MPs may be smaller than their size. Since the MP mass profile
is centrally concentrated, we adopt the modified definition \begin{equation}
b_{\min}\!=\!\min(0.1R_{p},GM_{p}/v^{2})\,,\label{e:bmin}\end{equation}
 where $v^{2}\!=\! GM(<\! r)/r$. This results in $\log\Lambda\!\sim\!6$
for MPs , less than the typical value for relaxation by stars. Nevertheless,
the error introduced by neglecting encounters with $b\!<\! b_{\min}$
is still not very large because penetrating encounters are much less
efficient. However, the assumption of multiple non-coherent encounters
with MPs over one orbital period is not necessarily justified because
of their small number density.

To address this, we modify the treatment of the empty loss-cone regime
(the contribution to the event rate from regions where the loss-cone
is already filled by stellar relaxation can not be increased by MPs,
see \S\ref{s:interact}). We define rare encounters as those with
impact parameters $b\!\le\! b_{1}$, where $b_{1}$ is defined by
\begin{equation}
P\int_{b_{\min}}^{b_{1}}\mathrm{d}b(\mathrm{d}\Gamma/\mathrm{d}b)\!=\!1\,.\label{e:b1}\end{equation}
 The differential rate is estimated simply by $(\mathrm{d}\Gamma/\mathrm{d}b)\!=\! N_{p}v2\pi b$.
When $P\int_{b_{1}}^{b_{\max}}\mathrm{d}b(\mathrm{d}\Gamma/\mathrm{d}b)\!>\!1$,
with $b_{max}\!=\! r$, all encounters with $b>b_{1}$ are defined
as frequent encounters that occur more than once per orbit, and add
non-coherently%
\footnote{In the marginal cases of $P\int_{b_{\min}}^{b_{\max}}\mathrm{d}b(\mathrm{d}\Gamma/\mathrm{d}b)\!<\!1$
or $P\int_{b_{1}}^{b_{\max}}\mathrm{d}b(\mathrm{d}\Gamma/\mathrm{d}b)\!<\!1$,
all encounters are considered rare. %
}. Note that even when $P\int_{b_{1}}^{b_{\max}}\mathrm{d}b(\mathrm{d}\Gamma/\mathrm{d}b)\!>\!1$
for all MPs, perturbations by rare, very massive MPs may still occur
less than once per orbit. Our treatment is approximate. A complete
statistical treatment of this situation lies beyond the scope of this
study.

When the typical number of encounters per orbit is less than one,
the fractional contributions from different individual encounters,
$\delta J$, should be averaged coherently ($\Delta J\!\propto\! t$),
subject to the limit that each encounter can at most fill the loss-cone.
The loss-cone limited change in angular momentum per orbit due to
rare encounters is therefore

\begin{equation}
j_{R}^{2}(E)=\left[P\!\int_{b_{\min}}^{b_{1}}\mathrm{d}b\frac{\mathrm{d}\Gamma}{\mathrm{d}b}\min\left(\delta J,J_{lc}\right)\right]^{2}\,.\label{e:jR}\end{equation}
 In contrast, frequent uncorrelated collisions add up non-coherently
($\Delta J\!\propto\!\sqrt{t}$), and it is only their final value
that is limited by the loss-cone (individual steps $\delta J$ may
exceed $J_{lc}$, but can then be partially cancelled by opposite
steps during the same orbit). The loss-cone limited change in angular
momentum per orbit due to frequent encounters is therefore

\begin{equation}
j_{F}^{2}(E)=\min\left[\frac{1}{\log(J_{c}/J_{lc})}P\!\int_{b_{1}}^{b_{\max}}\!\mathrm{d}b\frac{\mathrm{d}\Gamma}{\mathrm{d}b}\delta J^{2},J_{lc}^{2}\right]\,.\label{e:jF}\end{equation}
 The total loss-cone limited angular momentum change per orbit is
then approximated by \begin{equation}
j^{2}=\min\left(j_{R}^{2}+j_{F}^{2},J_{lc}^{2}\right)\,,\label{e:jtot}\end{equation}
 and the differential event rate is calculated by Eq. (\ref{e:dGdE}),
$\mathrm{d}\Gamma/\mathrm{d}E\!=\!\left[j^{2}(E)/J_{c}^{2}(E)\right]N_{\star}(E)/P(E)$.

The contribution of rare encounters is evaluated in the impulse approximation
by setting $\delta J\!\sim\! GM_{p}r/bv$ in Eq. (\ref{e:jR}). We
find that the this contribution by GC MPs (\S\ref{s:GC_MPs}) is
generally small. Frequent encounters are the regime usually assumed
in the Fokker-Planck treatment of the loss-cone problem (e.g. \citealt{lig+77}).
To evaluate the contribution of frequent encounters, we do not calculate
$\delta J$ directly, but instead calculate the sub-expression $I=P\int_{b_{1}}^{b_{\max}}\mathrm{d}b(\mathrm{d}\Gamma/\mathrm{d}b)\delta J^{2}$
in Eq. (\ref{e:jF}) in terms of the $b$-averaged diffusion coefficient
$D(v_{t}^{2})$, after averaging over the orbit between the periapse
$r_{p}$ and apoapse $r_{a}$ and averaging over the perturber mass
function (this is essentially equivalent to the definition of $J_{D}$
in terms of $t_{r}$, \S\ref{ss:noncoherent}),

\begin{eqnarray}
I & = & \int\mathrm{d}M_{p}\left(2\int_{r_{p}}^{r_{a}}\frac{r^{2}D(\Delta v_{t}^{2})}{v_{r}}\mathrm{d}r\right)\nonumber \\
 & \simeq & \int\mathrm{d}M_{p}\left(2\int_{0}^{2r}\frac{r^{2}D(\Delta v_{\perp}^{2})}{v}\mathrm{d}r\right)\,.\label{e:IF}\end{eqnarray}
 The assumptions involved in the last approximate term (\citealt{mag+99})
are that the star is on a nearly radial orbit ($v_{r}\!\rightarrow\! v$,
$r_{p}\!\rightarrow\!0$, $r_{a}\!\rightarrow\!2r$) and that $D(v_{t}^{2}$)
(the diffusion coefficient of the transverse velocity relative to
the MBH) can be approximated by $D(\Delta v_{\perp}^{2})$ (the diffusion
coefficient of the transverse velocity relative to the stellar velocity
$\mathbf{v}$), given explicitly by (\citet{bin+87}, Eq. 8-68) \begin{equation}
D(\Delta v_{\perp}^{2})=\frac{8\pi G^{2}(\mathrm{d}N_{p}/\mathrm{d}M_{p})M_{p}^{2}\ln\Lambda}{v}K\left(\frac{v}{\sqrt{2}\sigma}\right)\,,\label{e:delta_v}\end{equation}
 where $K(x)\!\equiv\!\mathrm{erf}(x)(1-1\left/2x^{2}\right.)+\exp(-x^{2})\left/\sqrt{\pi}x\right.$
and where a spatially homogeneous distribution of MPs with a Maxwellian
velocity distribution of rms 1D velocity $\sigma$ was assumed.

\begin{figure}
\plotone{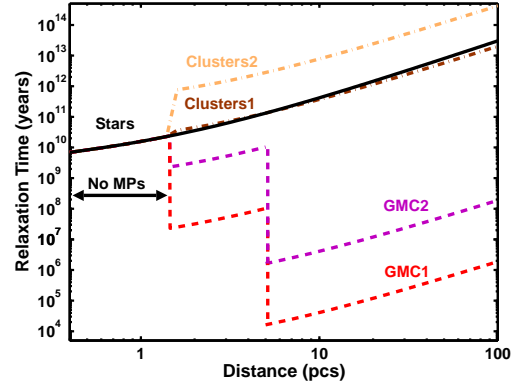}

\caption{\label{f:tr}Relaxation time as function of distance from the MBH,
for stars (solid line) and for each of the 4 MP models separately,
as listed in table \ref{t:models}: clusters (dashed-dotted lines),
GMCs (dashed lines). The discontinuities are artifacts of the assumed
sharp spatial cutoffs on the MP distributions. 2-body stellar processes
dominate close to the MBH, where no MPs are observed to exist. However,
at larger distances massive clumps (at $1.5<r<5$ pcs) and GMCs (at
$5<r<100$) are much more important. }
\end{figure}

To summarize, the event rates are calculated as follows. For each
perturber model (table \ref{t:models}), we integrate over the stellar
distribution ($N_{\star}\!=\!1.2\times10^{6}(r/0.4\,\mathrm{pc)^{-2}}$,
for $r\!>\!0.4\,\mathrm{pc}$ and $\Ms\!=\!1\,\Mo$) in terms of $r$,
using $N_{\star}$ to derive the appropriate density of perturbed
objects (single stars \S\ref{ss:single} or binaries \S\ref{ss:binary}).
At each $r$ we calculate $b_{\min}$ (Eq. \ref{e:bmin}), $b_{1}$
(Eq. \ref{e:b1}), $j_{R}$ (Eq. \ref{e:jR}) and $j_{F}$ (Eq. \ref{e:jF}).
The integral $I$ (Eq. \ref{e:IF}) is evaluated by taking $v^{2}\!\rightarrow\! GM(<r)/r$
and correcting approximately for the difference relative to the exact
calculation. We use $j$ (Eq. \ref{e:jtot}) to calculate the differential
event rate $\mathrm{d}\Gamma/\mathrm{d}r$ (Eq. \ref{e:dGdE}), with
the \emph{ansatz} $E\!\rightarrow GM(<r)\left/2a\right.$ where $a\!\equiv\!(4/5)r$
is an effective semi-major axis, which is motivated by the fact that
for a Keplerian isothermal eccentricity distribution, $\left\langle r\right\rangle \!=\! a(1+\left\langle e^{2}\right\rangle /2)\!=\!(5/4)a$.
The total event rate is calculated by carrying the integration over
$r$ in the region where MPs exist, between $r_{\mathrm{MP}}$ and
$r_{\mathrm{out}}$ (\S\ref{s:GC_MPs}).

\section{Massive perturbers in the Galactic Center}

\label{s:GC_MPs}

MPs can dominate relaxation only when they are massive enough to compensate
for their small space densities. Here we consider only MPs with masses
$M_{p}\!\ge\!10^{2}M_{\odot}$. Such MPs could be molecular clouds
of different masses, in particular giant molecular clouds (GMCs),
open or globular stellar clusters, and perhaps also IMBHs. As discussed
below, observations of the Galaxy reveal enough MPs to dominate relaxation
in the central 100 pc. We adopt here a conservative approach, and
include in our modeling only those MPs that are directly observed
in the Galaxy, namely GMCs and young clusters. We discuss briefly
theoretical predictions for two other classes of MPs: dynamically
evolved {}``submerged'' clusters and IMBHs, that could well be common
in galactic centers and contribute to efficient relaxation.

\begin{figure}
\plotone{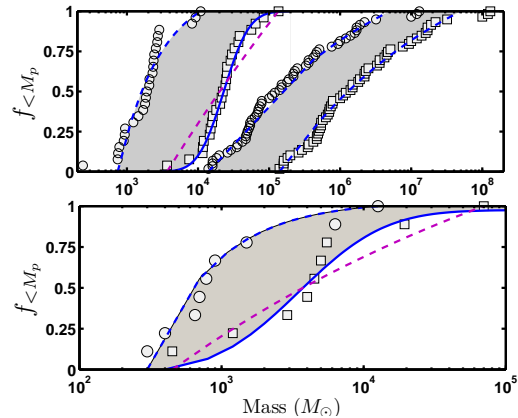}

\caption{\label{f:MP-MF}Cumulative mass functions of observed massive perturbers
in the GC (symbols) with best fit power-law differential distributions,
$dN/dM\propto M^{-\beta}$(dashed lines) or lognormal distribution
with log mean $\mu$and log standard deviation $\sigma$ (solid lines).
Top panel: lower (circles) and upper (squares) estimates for the masses
of observed molecular clumps \citep{chr+05} (left) and GMCs \citep{oka+01}
(right) in the GC . The best fit power-law indices for the GMCs are
$\beta_{up}\!=\beta_{low}\!=1.2$ and for the clumps $\beta_{up}\!=\!1.1$
and $\beta_{low}\!=\!1.7$. For the upper estimates of the clumps
masses a better fit is found by a log normal distribution with $\mu=8.2,\,\sigma=0.65$.
Bottom panel: likewise for GC clusters \citep{fig+99,fig04,bor+05}
with $\beta_{up}\!=\!1.3$ and $\beta_{low}\!=\!1.9$, and $\mu=10,\,\sigma=1.1$
for the upper masses estimate. }
\end{figure}

The dynamically dominant MPs are GMCs. Emission line surveys of the
central $\sim\!100$ pc reveal $\sim\!100$ GMCs with estimated masses
in the range $10^{4}$--$5\!\times\!10^{7}\,\Mo$ and sizes of $R_{p}\!\sim\!\mathrm{few\, pc}$
\citep{miy+00,oka+01,gus+04}. We selected individual, reliably identified
GMCs in the central $0.7^{\circ}$ of the Galaxy ($\sim\!100$ pc
of the GC), from the sample observed by \citet{oka+01}. Figure \ref{f:MP-MF}
shows the empirical GMC mass function using their virial mass estimates
as an upper mass limit and adopting a lower limit 10 times smaller,
following \citet{miy+00} who found that LTE mass estimates are typically
an order of magnitude lower than the virial ones. Note that the more
recent GMC CO1-0 molecular line observations by \citet{oka+01}, which
we use here, indicate a more massive and flatter mass function than
that derived for their earlier CS1-0 molecular line observations \citep{miy+00}.
This is probably due to the higher sensitivity of the CO1-0 line to
lower-density molecular gas (M. Tsuboi, priv. comm.).

Inside the inner $5\,\mathrm{pc}$ of the GC (a volume smaller or
comparable to that of a GMC or a stellar cluster), the most massive
local structures are the molecular gas clumps observed in the circumnuclear
gaseous disk (CND) and its associated spiral-like structures \citep{gen+85,chr+05}.
Their size are $\sim\!0.25$ pc and their masses are estimated to
be in the range $10^{3}$--$10^{5}\,\Mo$, where the lower estimates
are based on the assumption of optically thin HCN(1-0) line emission
and the upper estimates are based on the optically thick assumption,
which also coincides with the virial estimates. 

It is possible to obtain a model-independent estimate of the effect
of the \emph{observed} MPs on the relaxation time by the directly
derived value $\mu_{2}^{\mathrm{obs}}\!=\!\sum_{i}M_{p,i}^{2}/N_{\star}M_{\star}^{2}$,
which is listed in table \ref{t:models}. The observed GMC masses
show that $\mu_{2}^{\mathrm{obs}}\!\sim\!2\times10^{6-8}$ on the
100 pc scale, and $\mu_{2}^{\mathrm{obs}}\!\sim\!6\times10^{1-3}$
on the 5 pc scale, a clear indication that MPs dominate the relaxation
on all relevant lengthscales. where MPs exist. For the purpose of
our numeric calculations below, it is convenient to describe the differential
mass function analytically. Here we adopt a power-law $dN_{p}/\mathrm{d}M_{p}\!\propto\! M_{p}^{-\beta}$
parameterization (or a lognormal probability distribution function
(LNPDF) in the appropriate cases $dN_{p}/\mathrm{d}M_{p}\!\propto\! LNPDF(\mu,\,\sigma)$,
where$\mu$ is the log mean and $\sigma$is the log standard deviation).
Figure \ref{f:MP-MF} shows that the power law distibution is a good
fit for the GMCs mass function \citep{miy+00}, and for the lower
estimates of the clusters and gas clumps. For the upper estimates
of the clusters and the molecular gas clumps the MFs were found to
be better fit by a lognormal distribution, which we used in these
cases. It should be emphasized that our results and conclusions are
determined primarily by the large values of $\mu_{2}$, and not by
the detailed form of the mass function. We repeated our calculations
with several alternative distributions and found qualitatively similar
results. The assumed high mass cutoff of the MF is important, as it
determines the magnitude of $\mu_{2}$. Figure \ref{f:MP-MF} shows
that our models do not extrapolate beyond the maximum observed MP
masses (and even fall below them for GMCs). 

We obtained best fit power indices $\beta$ and LNPDF indices $\mu$and
$\sigma$ for the lower and upper estimates. The cumulative mass functions
and best fits are shown in Fig. (\ref{f:MP-MF}) and listed in table
\ref{t:MPs}. We find $\beta\!=\!1.2$ for both the lower mass range
($1.4\times10^{4}\le M_{p}\le5\times10^{6}\, M_{\odot}$) and the
upper GMC mass range ($1.4\times10^{5}\le M_{p}\le5\times10^{7}\, M_{\odot}$).
The clump mass function has $\beta\!\simeq\!1.1$ for the lower mass
range ($2.4\times10^{2}\le M_{p}\le1.1\times10^{4}\, M_{\odot}$)
and $\mu=10.04,\,\sigma=0.65$ ($\beta\!\simeq\!1.7$ for the best
power law fit) for the upper mass range ($3.6\times10^{3}\le M_{p}\le1.35\times10^{5}\, M_{\odot}$)
(Fig. \ref{f:MP-MF} and table \ref{t:MPs}). The space density of
such clumps falls rapidly inside the inner $\sim\!1.5$.

Stellar clusters are another class of MPs, which are of minor dynamical
significance in this context. About $10$ young stellar clusters with
masses in the range $10^{2}$--$10^{5}\,\Mo$ and sizes of order $R_{p}\!\sim\!1$
pc were observationally identified \citep{fig+99,fig+02,mai+04,fig+04,bor+05}.
Again, we fit the lower and upper mass estimates of these clusters
(with mass ranges $3\times10^{2}-1.3\times10^{4}\, M_{\odot}$ and
$4.5\times10^{2}-7\times10^{4}\, M_{\odot}$, respectively) with a
power-law mass function of a LNPDF, and find $\beta\!\simeq\!1.3$
and $\mu\simeq8.65,\,\sigma=1.1$ ($\beta=1.9$ for the best power
law fit; see Fig. \ref{f:MP-MF} and table \ref{t:MPs}). It is interesting
to note in passing that both the GMCs and gas clumps and the clusters
have bery similar mass functions distributions, but the clusters have
a lower mass range, as might be expected if these GMCs are the progenitors
of young GC clusters, similar to the relation observed between galactic
disk clusters and GMCs \citep{lad+03}.

Based on the current observations of the 9 confirmed clusters in the
GC, we find that they are dynamically insignificant compared to the
GMCs. Note however that their contribution to the relaxation is comparable
to that of the stars, even when the lower mass estimate is assumed.
When the upper mass estimate is assumed, clusters shorten the relaxation
time by a factor of 10 (Fig. \ref{f:tr} and table \ref{t:models}).
Clusters could play a larger role in relaxation if there are many
more yet undiscovered ones in the GC. Based on dynamical simulations,
\citet{por+06} suggest that there may be ${O}(100)$ evolved young
clusters in the region, undetected against the field stars because
of their low surface density. However, \citet{fig+02} argue that
these massive clusters, it they exist, should have been observed.
Two young GC cluster candidates were recently discovered by \citet{bor+06},
one of which may be more massive than $10^{4}\,\Mo$. It is thus possible
that the GC harbors some additional undetected massive clusters, although
probably not as many as suggested by \citet{por+06}. Such young clusters
may grow IMBHs by runaway stellar collisions (e.g. \citealt{por+04,fre+04}),
which would then sink to the center by dynamical friction, dragging
with them the cluster core. \citet{por+06} estimate that many IMBHs
may have migrated to the central 10 pc of GC in this way, however
there is as yet no observational evidence supporting this idea.

The MBH's dynamical influence extends up to a radius $r_{h}$ containing
${\cal O}(\Mbh$) mass in stars, which in the GC falls in the range
$2$--$4$ pc, depending on the operative definition of $r_{h}$ and
the uncertainties in the value of $\Mbh$ and of the density of the
surrounding stellar cluster (see \citealt{ale05}, sec 3.1.2). Thus
most Galactic MPs, and in particular the massive ones, lie outside
$r_{h}$. Table (\ref{t:MPs}) summarizes the estimated properties
of the observed MPs .

\begin{table*}

\caption{\label{t:MPs}Abundances of observed massive perturbers in the Galactic
Center}

\begin{centering}\begin{tabular}{lcccccccl}
\hline 
MP type&
$r$$^{a}$ (pc)&
$N_{p}$ &
$M_{p}$ ($M_{\odot}$) &
$\beta$&
$(\mu,\,\sigma)$&
$\left\langle M_{p}^{2}\right\rangle ^{1/2}\,(\Mo$)&
$R_{p}$ (pc)&
References\tabularnewline
\hline 
Observed GMCs&
$<\!100$ &
\textasciitilde{}100 &
$10^{4}-10^{8}$&
$1.2$&
&
$3\times10^{6}\,-\!3\times10^{7}$&
5&
\citet{oka+01,gus+04}\tabularnewline
Observed clusters&
$<\!100$&
\textasciitilde{}10&
$10^{2}-10^{5}$&
$1.3$&
$(8.2,\,1.1)$&
$4.8\,\times\,10^{3}\,-\,2.4\!\times\!10^{4}$&
1&
\citet{fig+99,fig+02,fig04}\tabularnewline
&
&
&
&
&
&
&
&
\citet{mai+04,bor+05}\tabularnewline
Observed clumps&
$1.5-3$&
\textasciitilde{}25&
$10^{2}-10^{5}$&
$1.1-1.7$&
$(10,\,0.65)$&
$3.7\,\times\,10^{3}\,-\,4.1\!\times\!10^{4}$&
0.25&
\citet{gen+85,chr+05}\tabularnewline
\hline 
\multicolumn{5}{l}{$^{a}$Projected distance range enclosing observed MPs.}&
\tabularnewline
\hline
\end{tabular}\par\end{centering}
\end{table*}

The observed MP species vary in their spatial distributions and mass
functions, which are not smooth or regular, and their distribution
constantly changes as their orbits decay by dynamical friction and
new MPs are formed and destroyed. It is thus likely that the presently
observed MPs are but one realization of a much smoother and regular
(well-mixed) underlying distribution. Consequently, we construct several
simplified MP models for our numeric calculations (table \ref{t:models})
that are based on the observed properties of the MPs (\S\ref{s:interact}),
and our best fits for their mass functions. The simplifications involve
the following assumptions. (i) A smooth, spherically symmetric MP
number density distribution, $N_{p}(r)\!\propto r^{-2}$ between the
cutoffs $\rMP\!<\! r\!<\! r_{\mathrm{out}}$ ($r_{\mathrm{out}}$
does not play an important role, see below), with a random velocity
field. (ii) A single or broken power-law MP mass functions, $\mathrm{d}N_{p}/\mathrm{d}M_{p}\!\propto\! M_{p}^{-\beta}$
(or a lognormal distribution in the appropriate cases). (iii) A single
mass stellar population of $1\,\Mo$ stars with a number density distribution
$\Ns(r)\!\propto r^{-2}$ outside the inner 1.5 pc (i.e. a constant
MP to star ratio). (iv) Mutually exclusive perturber types (i.e.,
a single type of perturber is assumed to dominate relaxation, as indicated
by the detailed calculations presented in Fig. \ref{f:tr}).

The five MP models are detailed in table \ref{t:models}: Stars, Clusters1,
Clusters2, GMC1 and GMC2, represent respectively the case of relaxation
by stars only, by heavy and light stellar clusters, and by heavy GMCs
and light GMCs. Table \ref{t:models} lists $\mu_{2}$, the ratio
of the 2nd moment of the various MP mass distributions to that of
the stars.

\begin{table}

\caption{\label{t:models}Massive perturbers models }

\begin{centering}\begin{tabular}{ccccrrcc}
\hline 
\multicolumn{1}{c}{Model}&
$r$ (pc)$\,{}^{a}$&
$N_{p}\,^{b}$&
$M_{p}$ ($M_{\odot}$)&
$\beta$&
$\mu,\,\sigma$&
$R_{p}$ (pc)&
$\mu_{2}\,^{c}$\tabularnewline
\hline 
GMC1&
5--100 &
$100$&
$10^{5}\!-\!5\times10^{7}$&
$1.2$&
--&
5&
$2\!\times\!10^{8}$\tabularnewline
&
1.5--5&
$30$&
$3\times10^{3}\!-\!10^{5}$&
$1.1$&
$10.04,\,0.65$&
&
$5500$\tabularnewline
GMC2&
5--100 &
$100$&
$10^{4}\!-\!5\times10^{6}$&
$1.2$&
--&
5&
$2\!\times\!10^{6}$\tabularnewline
&
1.5--5&
$30$&
$7\times10^{2}\!-\!10^{4}$&
$1.7$&
--&
&
$44$\tabularnewline
Clusters1&
1.5--100 &
$10$&
$5\times10^{2}\!-\!7\times10^{4}$&
$1.3$&
--&
1&
$27$\tabularnewline
Clusters2&
1.5--100 &
$10$&
$3\times10^{2}\!-\!10^{4}$&
$1.9$&
$8.16,\,1.1$&
1&
$1$\tabularnewline
Stars&
5--100 &
$2\!\times\!10^{8}$&
$1$&
---&
--&
$\sim\!0$&
1\tabularnewline
Stars&
1.5--5 &
$8\!\times\!10^{6}$&
$1$&
---&
--&
$\sim\!0$&
1\tabularnewline
\hline 
\multicolumn{5}{l}{$^{a}$Distance range enclosing observed MPs.}&
\tabularnewline
\multicolumn{5}{l}{{\footnotesize \rule{0em}{1.5em}$^{b}$} $N_{p}(r)\!\propto\! r^{-2}$
in the given range is assumed.}&
\tabularnewline
\multicolumn{5}{l}{{\footnotesize \rule{0em}{1.5em}$^{c}$} $\mu_{2}^{\mathrm{obs}}\!=\!\sum_{i}M_{p,i}^{2}/N_{\star}M_{\star}^{2}$,
where $M_{p,i}$ is the observed MP's mass. }&
\tabularnewline
\hline
\end{tabular}\par\end{centering}
\end{table}

\section{Massive perturber-driven interactions with a MBH}

\label{s:interact}

\begin{figure}
\begin{tabular}{c}
\plotone{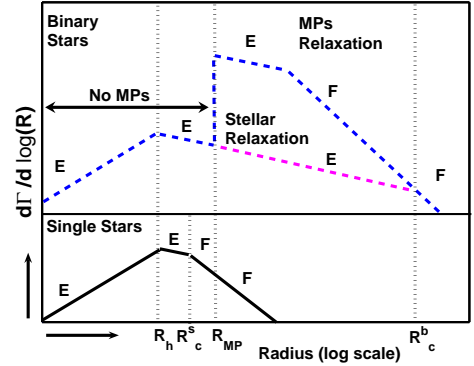} \tabularnewline
\end{tabular}

\caption{\label{f:bin} A schematic representation of the local contribution
to the tidal disruption rate for a single component MP model (e.g.
GMC2 without the central clumps, so that $\rMP\!=\!5$ pc). Top: the
disruption rate of binaries (a constant binary fraction is assumed
here for simplicity) due to stellar relaxation (bottom dashed line)
and MPs (top dashed line). Bottom: the disruption rate of single stars
for both perturber models. Empty and full loss-cone regimes are denoted
by {}``E'' and {}``F'', respectively. The initial orbital energy
of the disrupted objects is expressed by the corresponding radius
of a circular orbit, $r$. For $r\!<\! r_{h}$, stellar density and
velocity distributions of $\Ns(r)\!\propto\! r^{-3/2}$ and $\sigma(r)\!\propto r^{-1/2}$
are assumed; for $r\!>\! r_{h}$, $\Ns(r)\!\propto\! r^{-2}$ and
$\sigma(r)\!=\!\mathrm{const}.$ The transition from the empty to
full loss-cone regime (for stellar relaxation) occurs at the critical
radius $r_{c}^{s}\!<\!\rMP$ for single stars and at $r_{c}^{b}\!\gg\!\rMP$
for binaries. }
\end{figure}

The maximal differential loss-cone refilling rate, which is also the
close encounters event rate, $\mathrm{d}\Gamma/\mathrm{d}E$, is reached
when relaxation is efficient enough to completely refill the loss
cone during one orbit (Eq. \ref{e:dGdE}). Further decrease in the
relaxation time does not affect the event rate at that energy. MPs
can therefore increase the differential event rate over that predicted
by stellar relaxation, only at high enough energies, $E\!>\! E_{c}$
(equivalently, small enough typical radii, $r\!<\! r_{c}$), where
slow stellar relaxation fails to refill the empty loss-cone. The extent
of the empty loss-cone region increases with the maximal periapse
$q$, which in turn depends on the close encounter process of interest.
For example, the tidal disruption of an object of mass $M$ and size
$R$ occurs when $q\!<\! r_{t}$, the tidal disruption radius, \begin{equation}
r_{t}\simeq R\left(\Mbh/M\right)^{1/3}\,.\label{e:rt}\end{equation}
 This approximate disruption criterion applies both for single stars
($M\!=\!\Ms$, $R\!=\!\Rs$) and for binaries, where $M$ is the combined
mass of the binary components and $R$ is the binary's semi-major
axis, $a$. Stellar radii are usually much smaller than typical binary
separations, but stellar masses are only $\sim\!2$ times smaller
than binary masses. Binaries are therefore disrupted on larger scales
than single stars. In the GC this translates to an empty (stellar
relaxation) loss-cone region extending out to $r_{c}^{s}\!\sim\!3$
pc for single stars and out to $r_{c}^{b}\!>\!100$ pc for binaries.
In the GC $\rMP\!\lesssim\! r_{c}^{s}\!\ll\! r_{c}^{b}$ , and so
MPs are expected to increase the binary disruption rate by orders
of magnitude, but increase the single star disruption rate only by
a small factor. This is depicted qualitatively in figure (\ref{f:bin}),
which shows the local rates ($\mathrm{d}\Gamma/\mathrm{d}\log r$)
for disruption of single stars and binaries due to stellar relaxation
or relaxation by a simplified one component MP model.

Figure (\ref{f:bin}) also shows that the MP-induced disruption rate
is dominated by binaries originating near the \emph{inner} cutoff
$\rMP$ (in the following discussion the initial orbital energy of
the disrupted objects is expressed by the corresponding radius of
a circular orbit, $r\!\sim\! GM(<\! r)/2E$). This is qualitatively
different from the usual case of stellar tidal disruption induced
by stellar relaxation, which mainly occurs inside $r_{c}^{s}$ and
is dominated by stars originating near the \emph{outer} limit $\min(r_{c}^{s},r_{h})$;
usually $r_{c}^{s}\!\sim\! r_{h}$ \citep{lig+77}. The difference
can be understood by considering the $r$-dependence of $\mathrm{d}\Gamma\left/\mathrm{d}\log r\right.$.
Neglecting logarithmic terms, the empty and full local loss-cone rates
are, respectively (Eqs. \ref{e:Gempty}, \ref{e:Gfull})\begin{eqnarray}
\frac{\mathrm{d}\Gamma_{e}}{\mathrm{d}\log r} & \sim & \frac{N_{\star}(<\! r)}{t_{r}(r)}\,,\label{e:Gapprox}\\
\frac{\mathrm{d}\Gamma_{f}}{\mathrm{d}\log r} & \sim & \left[\frac{\Mbh}{\Mbh+\Ms(<\! r)}\right]\left(\frac{q}{r}\right)\frac{\Ns(<\! r)}{P(r)}\,,\end{eqnarray}
 where $\Ns(<r)$ is the number of stars enclosed within $r$.

Inside $r_{h}$, the potential is dominated by the MBH ($\Ms(<\! r)/M_{\bullet}\!\ll\!1$),
$\Ns\!\propto r^{-\alpha_{1}}$ with $\alpha_{1}<2$ for most dynamical
scenarios (e.g. \citealt{bah+77,you80}) and the velocity dispersion
is Keplerian, $\sigma\!\propto\! r^{-1/2}$. The orbital period scales
as $P\!\propto\! r^{3/2}$ and $t_{r}\sim(\Mbh/\Ms)^{2}P/[(\log\Ns)\Ns]$.
The local disruption rates are then $\left.\mathrm{d}\Gamma_{e}\left/\mathrm{d}\log r\right.\right|_{r<r_{h}}\!\propto\! r^{9/2-2\alpha_{1}}$
and $\left.\mathrm{d}\Gamma_{f}\left/\mathrm{d}\log r\right.\right|_{r<r_{h}}\!\propto\! r^{1/2-\alpha_{1}}$.
For plausible values of $1/2\!<\!\alpha_{1}\!<\!9/4$, $\Gamma_{e}$
increases with $r$ whereas $\Gamma_{f}$ decreases with $r$, so
the rate is dominated by stars near $r_{c}$ \citep{lig+77}. Outside
$r_{h}$, the stellar distribution is usually assumed to be near isothermal,
$\Ns\!\propto r^{-\alpha_{2}}$ with $\alpha_{2}\sim\!2$ and a velocity
dispersion $\sigma\!\sim\!\mathrm{const}$, and the potential is dominated
by the stars ($\Mbh\!\ll\!\Ms(<\! r)\propto r^{3-\alpha_{2}}$). The
orbital period scales as $P\!\propto\! r^{\alpha_{2}/2}$ and $t_{r}\!\sim\!\Ns P/\log\Ns$.
The local disruption rates are then $\left.\mathrm{d}\Gamma_{e}\left/\mathrm{d}\log r\right.\right|_{r>r_{h}}\!\propto\! r^{-\alpha_{2}/2}\!\sim\! r^{-1}$
and $\left.\mathrm{d}\Gamma_{f}\left/\mathrm{d}\log r\right.\right|_{r>r_{h}}\!\propto\! r^{-1-\alpha_{2}/2}\sim r^{-2}$.
For $\alpha_{2}\!\sim\!2$ both rates increase toward small radii.
Since for most Galactic MP types $\rMP\!>\! r_{h}$, the disruption
rate is dominated by stars near $r_{\mathrm{MP}}$. For example, when
the loss-cone is empty, $\sim\!50\%$ of the total rate is due to
MPs at $r\!<\!2\rMP$; when the loss-cone is full, $\sim\!75\%$ of
the total rate is due to MPs at $r\!<\!2\rMP$.

\subsection{Interactions with single stars}

\label{ss:single}

GMCs, gas clumps and clusters in the GC are abundant only beyond the
central $r_{\mathrm{MP}}\!\sim\!1.5$ pc, whereas the empty loss-cone
regime for tidal disruption of single stars extends only out to $r_{c}^{s}\!\sim\!3$
pc. For inspiral processes such as GW emission, $r_{c}$ is $\sim\!100$
times smaller still \citep{hop+05,hop+06b}. The effect of such MPs
on close encounter events involving single stars is thus suppressed
(weaker tidal effects by MPs at $r\!>\! r_{c}^{s}$ are not considered
here). This is contrary to the suggestion of \citet{zha+02}, who
assumed that the effect of MPs fully extends to the empty loss-cone
regime. We find that the enhancement of MPs over stellar relaxation
to the single stars disruption rate is small, less than a factor of
$3$, and is due to stars scattered by gas clumps in the small empty-loss
cone region between $\rMP\!\sim\!1.5\,\mathrm{pc}$ and $r_{c}^{s}\!\sim\!3\,\mathrm{pc}$.
A possible exception to this conclusion is the hypothesized population
of IMBHs \citep{por+06}, not modeled here, whose distribution could
extend to the inner pc (e.g. the IMBH candidate in IRS 13, \citealt{mai+04},
but see \citealt{sch+05} and \citealt{pau+06}).

\subsection{Interactions with stellar binaries}

\label{ss:binary}

The empty loss cone region for binary-MBH interactions extends out
to $>\!100$ pc because of their large tidal radius. On these large
scales MPs are abundant enough to dominate the relaxation processes.
Here we focus on 3-body exchange interactions \citep{hil91,hil92,yuq+03},
which lead to the disruption of the binary, the energetic ejection
of one star, and the capture of the other star on a close orbit around
the MBH.

Various phenomena associated with such exchange interactions were
suggested and explored. \citet{hil88} and later \citet{yuq+03},
\citet{gua+05}, \citet{gin+06} and \citet{bro+06c}, studied hyper-velocity
stars ejected from the GC following tidal disruption by the MBH. \citet{gou+03}
suggested this mechanism to explain the origin of the young stars
near the Galactic MBH. \citet{mil+05} proposed that compact objects
captured following a binary disruption event will eventually be sources
of GWs from zero-eccentricity orbits, in contrast to high-eccentricity
sources typical of single star inspiral \citep{hop+05}.

The event rates estimated by these authors vary substantially. \citet{hil88}
assumed a full loss-cone and a fraction $f_{\mathrm{bin}}\!=\!0.02$
of the stars in binaries with small enough semi-major axis to produce
a high-velocity star ($a\!<\!0.1\,\mathrm{AU}$), and derived a 3-body
exchange rate of $\sim10^{-3}(f_{\mathrm{bin}}/0.02)\,\mathrm{yr^{-1}}$.
\citet{yuq+03} took into account the empty loss cone regime, and
argued for a higher fraction of relevant binaries ($f_{\mathrm{bin}}\!=\!0.04$
for binaries with $a\!<\!0.3\,\mathrm{AU}$ that can survive 0.8 pc
from the MBH), thereby obtaining a rate of $\sim2.5\times10^{-6}(f_{\mathrm{bin}}/0.04)\,\peryr$,
3 orders of magnitude smaller than that estimated by Hills. These
calculations assumed the same binary separation for all binaries and
a constant binary fraction at all distances from the MBH ( two possibilities
were considered, $a=0.3\, AU$ and $a=0.03\, AU$).

The binary fraction and typical binary semi-major axis depend on the
binary mass, and on the rate at which binaries evaporate by encounters
with other stars. This depends in turn on the stellar densities and
velocities, and therefore on the distance from the MBH. Here we take
these factors into account and estimate in detail the 3-body exchange
rate for MP-driven relaxation. The rate is proportional to the binary
fraction in the population, which is the product of the poorly-known
binary IMF in the GC and the survival probability against binary evaporation.

We assume for simplicity equal mass binaries, $M_{\mathrm{bin}}\!=\!2\Ms$,
since the observations indicate that moderate mass ratios dominate
the binary population \citep{duq+91,kob+06}. The evaporation timescale
at distance $r$ from the center for a binary of semi-major axis $a$
composed of two equal mass stars of mass $\Ms$ and lifetime $t_{\star}$
is (e.g. \citealt{bin+87}) \begin{equation}
t_{\mathrm{evap}}(\Ms,a,r)=\frac{M_{\mathrm{bin}}}{\left\langle \Ms\right\rangle }\frac{\sigma(r)}{16\sqrt{\pi}\rho(r)Ga\ln\Lambda_{\mathrm{bin}}}\,.\label{e:t_evap}\end{equation}
 The Coulomb factor for binary evaporation, $\Lambda_{\mathrm{bin}}\!=\! a\sigma^{2}/4G\left\langle \Ms\right\rangle $,
expresses the fact that the binary is only affected by close perturbations
at distances smaller than $\sim\! a/2$. The MPs considered here are
extended objects (table \ref{t:models}) and therefore do not affect
the binary evaporation timescale (IMBH MPs could be a possible exception).
Although binary evaporation is a stochastic process and the actual
time to evaporation differs from binary to binary, we expect a small
scatter in the evaporation rate, $\Delta t_{\mathrm{evap}}^{-1}/t_{\mathrm{evap}}^{-1}\!\ll\!1$,
because evaporation is a gradual process caused by numerous weak encounters.
Evaporation is thus better approximated as a fixed limit on the binary
lifetime, rather than as a Poisson process (where $\Delta t_{\mathrm{evap}}^{-1}/t_{\mathrm{evap}}^{-1}\!=\!1$).
The maximal binary lifetime is then $\tau\!=\!\min([t_{H},t_{\star}(\Ms),t_{\mathrm{evap}}(\Ms,a,r)]$,
where $t_{H}$ is the Hubble time, taken here to be the age of the
Galaxy. It is well established that the central $100$--$200\,\mathrm{pc}$
of the GC have undergone continuous star formation over the lifetime
of the Galaxy \citep{ser+96,fig+04}. Assuming a constant star formation
rate over time $t_{H}$, the differential binary distribution at time
$t$ is $\left.\mathrm{d}N_{\mathrm{bin}}/\mathrm{d}a\right|_{t}=f_{\mathrm{bin}}\left.\mathrm{d}f/\mathrm{d}a\right|_{0}\Gamma_{\star}\min(t,\tau)$,
where $\left.\mathrm{d}f/\mathrm{d}a\right|_{0}$ is the normalized
initial semi-major axis distribution, which can be observed in binaries
in low-density environments where $t_{\mathrm{evap}}\!\rightarrow\!\infty$,
and $\Gamma_{\star}$ is the single star formation rate, which is
normalized to the observed present day stellar density by setting
$t\!=\! t_{H}$ and taking $t_{\mathrm{evap}}\!\rightarrow\!\infty$
for singles, so that $\Gamma_{\star}\!=\!\Ns/\min(t_{H},t_{\star})$.
The present-day binary semi-major axis distribution is therefore \begin{eqnarray}
\left.\frac{\mathrm{d}N_{\mathrm{bin}}}{\mathrm{d}a}\right|_{t_{H}} & = & f_{\mathrm{bin}}(\Ms)\left.\frac{\mathrm{d}f}{\mathrm{d}a}\right|_{0}N_{\star}(r)\times\nonumber \\
 &  & \min\left\{ 1,\frac{t_{\mathrm{evap}}(\Ms,a,r)}{\min[t_{H},t_{\star}(\Ms)]}\right\} \,.\label{e:fhard}\end{eqnarray}

The capture probability and the semi-major axis distribution of captured
stars were estimated by simulations \citep{hil91,hil92,yuq+03}. Numeric
experiments indicate that between $0.5$--$1.0$ of the binaries that
approach the MBH within the tidal radius $r_{t}(a)$ (Eq. \ref{e:rt})
are disrupted. Here we adopt a disruption efficiency of $0.75$. The
harmonic mean semi-major axis for 3-body exchanges with equal mass
binaries was found to be \citep{hil91} \begin{equation}
\left\langle a_{1}\right\rangle \simeq0.56\left(\frac{M_{\bullet}}{M_{\mathrm{bin}}}\right)^{2/3}a\simeq\,0.56\left(\frac{M_{\bullet}}{M_{\mathrm{bin}}}\right)^{1/3}r_{t},\label{e:afinal}\end{equation}
 where $a$ is the semi-major axis of the infalling binary and $a_{1}$
that of the captured star (the MBH-star {}``binary''). Most values
of $a_{1}$ fall within a factor $2$ of the mean. This relation maps
the semi-major axis distribution of the infalling binaries to that
of the captured stars: the harder the binaries, the more tightly bound
the captured stars. The velocity at infinity of the ejected star (neglecting
the Galactic potential) is $v_{\mathrm{BH}}^{2}\!=\!2^{1/2}(GM_{\mathrm{bin}}/a)(M_{\bullet}/M_{\mathrm{bin}})^{1/3}\!\propto\! M_{\mathrm{bin}}^{2/3}/a$
(an equal mass binary with periapse at $r_{t}$ is assumed; \citealt{hil88}).
The harder the binary, the higher is $v_{\mathrm{BH}}$. The periapse
of the captured star is at $r_{t}$, and therefore its eccentricity
is very high \citep{hil91,hil92,mil+05}, $e\!=\!1-r_{t}/a_{1}\!\simeq1-1.8(M_{\mathrm{bin}}/M_{\bullet})^{1/3}\!\gtrsim\!0.98$
for values typical of the GC.

We now consider separately the implications of 3-body exchange interactions
of the MBH with old ($t_{\star}\!\gtrsim\! t_{H}$) binaries and massive
young ($t_{\star}\!<\!5\!\times10^{7}$ yr) binaries.

\subsubsection{Low mass binaries}

\label{sss:oldbin}

The properties of binaries in the inner GC are at present poorly determined.
The period distribution of Solar neighborhood binaries can be approximated
by a log normal distribution with a median period of 180 years ($a\!\sim\!40$
AU) \citep{duq+91}. The total binary fraction of these binaries is
estimated at $f_{\mathrm{bin}}\sim\!0.3$ \citep{lad06}. Adopting
these values for the GC, the total binary disruption rate by the MBH
can then be calculated, as described in \S\ref{s:outline}, by integrating
$\mathrm{d}N_{\mathrm{bin}}/\mathrm{d}a$ (Eqs. \ref{e:fhard}) over
the binary $a$ distribution and over the power-law stellar density
distribution of the GC up to 100 pc \citep{gen+03a}. Table (\ref{t:bin})
lists the capture rates for the different perturber models, assuming
a typical old equal-mass binary of $M_{\mathrm{bin}}\!=\!2\,\Mo$.

The old, low-mass binary disruption rate we derive for stellar relaxation
alone is $\sim5\times10^{-7}\,\mathrm{yr^{-1}}$, $\sim\!5$ times
lower, but still in broad agreement with the result of \citet{yuq+03}.
Their rate is somewhat higher because they assumed a constant binary
fraction and a constant semi-major axis for all binaries, even close
to the MBH, where these assumptions no longer hold.

MPs increase the binary disruption and high-velocity star ejection
rates by factors of $\sim\!10^{1-3}$ and effectively accelerate stellar
migration to the center. This can modify the stellar distribution
close to the MBH by introducing a {}``source term'' to the stellar
current into the MBH. Low-mass stars are at present too faint to be
directly observed in the GC. However, such a source term may have
observable consequences since it can increase the event rate of single
star processes such as tidal disruption, tidal heating and GW emission
from compact objects, in particular from compact objects on zero-eccentricity
orbits \citep{mil+05} (in contrast, GW from inspiraling single stars
occur on high-eccentricity orbits, \citealt{hop+05}). We calculated
numerical solutions of the Fokker-Planck equation for the stellar
distribution around the MBH with a captured stars source term. These
preliminary investigations (Perets, Hopman \& Alexander, in prep.)
confirm that the total accumulated mass of captured stars does not
exceed the dynamical constraints on the extended mass around the MBH
\citep{mou+05}, because 2-body relaxation and likely also resonant
relaxation \citep{rau+96,hop+06a} scatters enough of them into the
MBH or to wider orbits.

\begin{center}%
\begin{table*}

\caption{\label{t:bin}Total binary disruption rate and number of captured
young stars }

\begin{tabular}{>{\raggedright}p{0.6in}ccccc}
\hline 
Model&
\multicolumn{2}{c}{Disruption rate ($\mathrm{yr^{-1}}$) }&
Young Stars$^{a}$&
Young Stars$^{a}$ &
Young HVSs$^{b}$\tabularnewline
&
$r\!<\!0.04\,\mathrm{pc}$&
$r\!<\!0.4\,\mathrm{pc}$&
$r\!<\!0.04\,\mathrm{pc}$&
$0.04\,<\, r\!<\!0.4\,\mathrm{pc}$&
\tabularnewline
\hline 
GMC1&
$1.1\times10^{-4}$&
$3.2\times10^{-4}$&
$39.4$&
$5.2$&
342\tabularnewline
GMC2&
$2.8\times10^{-5}$&
$1.8\times10^{-4}$&
$6$&
$1.4$&
49\tabularnewline
Clusters1&
$6.4\times10^{-7}$&
$9.7\times10^{-7}$&
$0.19$&
$0.004$&
1.9\tabularnewline
Clusters2&
$2.6\times10^{-8}$ &
$4\times10^{-8}$ &
$0.01$ &
$10^{-4}$ &
0.1\tabularnewline
Stars&
$3.4\times10^{-7}$&
$5.3\times10^{-7}$&
$0.15$&
$0.003$&
1.3\tabularnewline
Observed&
?&
?&
$10-35^{c}$&
$?$&
$43\pm31^{d}$\tabularnewline
\hline 
\multicolumn{6}{p{3in}}{{\footnotesize $^{a}$Main sequence B stars with lifespan $t\!<\!5\!\times\!10^{7}\,\mathrm{yr}$
.}}\tabularnewline
\multicolumn{6}{l}{{\footnotesize $^{b}$Main sequence B stars with lifespan $t\!<\!4\!\times\!10^{8}\,\mathrm{yr}$. Notice that only ~20 percents of these ejected stars could be observed in regions covered by current surveys.   
.}}\tabularnewline
\multicolumn{6}{l}{{\footnotesize $^{c}\,$ $\sim\!10$ stars with derived $a\!\lesssim0.04\,\mathrm{pc}$.
$\gtrsim\!30$ stars are observed in the area.}}\tabularnewline
\multicolumn{6}{l}{$^{^{d}}${\footnotesize Estimated from the observed 5 HVSs, at distances between 20-120 pc from the GC \citep{bro+06a}}}\tabularnewline
\hline
\end{tabular}
\end{table*}
\par\end{center}

\subsubsection{Young massive binaries}

\label{sss:youngbin}MPs may be implicated in the puzzling presence
of a cluster of main sequence B-stars ($4\!\lesssim\!\Ms\!\lesssim\!15\,\Mo$)
in the inner $\sim\!1^{''}$ ($\sim\!0.04$ pc) of the GC. This so-called
{}``S-cluster'' is spatially, kinematically and spectroscopically
distinct from the young, more massive stars observed farther out,
on the $\sim\!0.05$--$0.5$ pc scale, which are thought to have formed
from the gravitational fragmentation of one or two gas disks \citep{lev+03,gen+03a,mil+04,pau+06}.
There is however still no satisfactory explanation for the existence
of the seemingly normal, young massive main sequence stars of the
S-cluster, so close to a MBH (see review of proposed models by \citealt{ale05};
also a recent model by \citealt{lev06}).

Here we revisit an idea proposed by \citet{gou+03}, that the S-stars
were captured near the MBH by 3-body exchange interactions with infalling
massive binaries. Originally, this exchange scenario lacked a plausible
source for the massive binaries. \citeauthor{gou+03} speculated that
they originated in an unusually dense and massive young cluster on
an almost radial infall trajectory, but concluded that such a finely-tuned
scenario seems unlikely. Furthermore, a massive cluster is expected
to leave a tidally stripped tail of massive stars beyond the central
$0.5$ pc \citep{kim+04,gur+05}, which are not observed \citep{pau+06}.
Alternatively, it must contain an unusually massive central IMBH to
hold it together against the tidal field of the GC \citep{han+03}.
However, such a massive IMBH is well beyond what is predicted by simulations
of IMBH formation by runaway collisions \citep{gur04,gur+05}, or
anticipated by extrapolating the $M/\sigma$ relation \citep{fer+00,geb+00}
to clusters.

MP-driven 3-body exchanges circumvent the problems of the cluster
infall scenario by directly bringing massive \emph{field} binaries
to $r_{t}$, without requiring massive clusters of unusual, perhaps
even impossible properties. The ongoing star formation in the central
$\sim\!100$ pc implies the presence of a large reservoir of massive
stars there, which are indeed observed in the central $\mathrm{few}\times10$
pc both in dense clusters and in the field \citep{fig+99,fig+03,mun+06}.
It is plausible that a high fraction of them are in binaries.

We model the binary population of the GC in the S-stars mass range
, $4\!\lesssim\! M_{\star}\!\lesssim\!15\, M_{\odot}$, by assuming
equal mass binaries that follow the single star mass function with
an initial binary fraction of $f_{\mathrm{bin}}\!\sim\!0.75$, as
observed elsewhere in the Galaxy (\citealt{lad06,kob+06}). Because
the stellar evolutionary lifespan of such stars is relatively short,
massive binaries are essentially unaffected by dynamical evaporation.
We assume star formation at a constant rate for 10 Gyr with a Miller-Scalo
IMF \citep{mil+79}, and use a stellar population synthesis code \citep{ste+03}
with the Geneva stellar evolution tracks \citep{sch+92a} to estimate
that the present day number fraction of stars in the S-star mass range
is $3.5\!\times\!10^{-4}$ (and less than 0.01 of that for $\Ms\!>\!15\,\Mo$
stars). Note that if star formation in the GC is biased toward massive
stars \citep{fig+03,st0+05}, this estimate should be revised upward.
We adopt the observed Solar neighborhood distribution of the semi-major
axis of massive binaries, which is log-normal with $\left\langle \log a\right\rangle =-0.7\pm$0.6
AU (i.e. 63\% of the binaries with $a\!=\!0.2_{-0.15}^{+0.60}$ AU;
91\% with $a\!=\!0.2_{-0.19}^{+2.96}$ AU) \citep{gar+80,kob+06}.
Massive binaries are thus typically harder than low-mass binaries,
and will be tidally disrupted (Eq. \ref{e:rt}) closer to the MBH
and leave a more tightly bound captured star.

We represent the massive binaries by one with equal mass stars in
the mid-range of the S-stars masses, with $M_{\mathrm{bin}}\!=\!15\,\Mo$
and $t_{\star}(7.5\,\Mo)\simeq5\!\times\!10^{7}\,\mathrm{yr}$, and
integrate over the stellar distribution and the binary $a$ distribution
as before, to obtain the rate of binary disruptions, $\Gamma$, the
mean number of captured massive stars in steady state, $N_{\star}\!=\!\Gamma t_{\star}$,
and their semi-major axis distribution (Eq. \ref{e:afinal}). Table
(\ref{t:bin}) compares the number of captured young stars in steady
state, for the different MP models, on the $r\!<\!0.04\,\mathrm{pc}$
scale (the S-cluster) and $0.04\!<\! r\!<\!0.4$ pc scale (the stellar
rings) with current observations \citep{eis+05,pau+06}.

\begin{figure}
\plotone{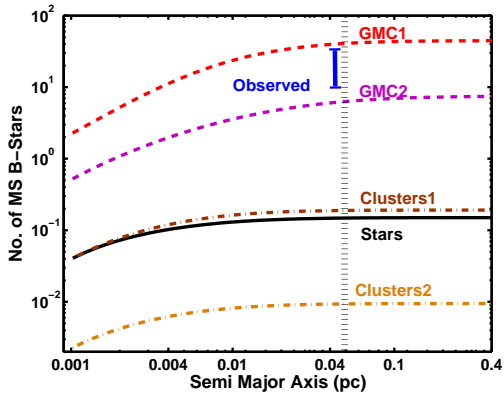}

\caption{\label{f:Scluster}Cumulative number of young B-stars in the GC as
predicted by the MP models and by stellar two-body relaxation (listed
in table 2).The vertical bar represents the total number of observed
young stars inside 0.04 pc (\citealt{eis+05}; S. Gillessen, priv.
comm.). The dotted vertical line marks the approximate maximal distance
in which captured B-stars are expected to exist. }
\end{figure}

The S-stars are found in the central $<\!0.04$ pc \citep{ghe+05,eis+05}.
If they were captured by binary disruptions, they must have originated
from massive binaries with $a\!\lesssim\!3.5$ AU. This is consistent
with semi-major axis distribution of massive binaries. The number
of captured massive stars falls rapidly beyond $0.04$ pc (table \ref{t:bin})
because wide massive binaries are rare. This capture model thus provides
a natural explanation for the central concentration of the S-cluster
(Fig \ref{f:Scluster}). The absence of more massive stars in the
S-cluster ($\Ms\!>\!15\,\Mo$, spectral type O\,V) is a statistical
reflection of their much smaller fraction in the binary population.
Figure (\ref{f:Scluster}) and table (\ref{t:bin}) compare the cumulative
semi-major axis distribution of captured B-stars, as predicted by
the different MP models, with the total number of young stars observed
in the inner 0.04 pc ($\sim\!35$ stars, \citealt{eis+05,ghe+05,pau+06}).
Of these, only $\sim\!10$ have full orbital solutions (in particular
$a$ and $e$) at present. For the others we assume the \emph{ansatz}
that $a$ is similar to the observed projected position. The numbers
predicted by the GMC-dominated MP models are consistent with the observations,
unlike the stellar relaxation model that falls short by two orders
of magnitude.

The binary capture model predicts that captured stars have very high
initial eccentricities. Most of the solved S-star orbits do have $e\!>\!0.9$,
but a couple have $e\!\sim\!0.3$--$0.4$ \citep{eis+05}. Normal,
non-coherent stellar relaxation is slow, even after taking into account
the decrease in $t_{r}$ toward the center due to mass segregation
\citep{hop+06b}. It is unlikely that it could have decreased the
eccentricity of these stars over their relatively short lifetimes.
However, the much faster process of resonant relaxation \citep{rau+96}
may be efficient enough to randomize the eccentricity of a fraction
of the stars, and could thus possibly explain the much larger observed
spread in eccentricities \citep{hop+06a}. Additional orbital solutions
and a better estimate of the efficiency of resonant relaxation in
the GC are required for more detailed comparisons between observations
and the MP model predictions.

\subsubsection{Hyper-velocity stars}

\label{sss:HVS}

Each captured star is associated with an ejected companion, which
in some cases is launched with a very high velocity. The one-to-one
correspondence between the number of captured S-stars and the number
of early-type hyper-velocity stars (HVSs) is thus a generic prediction
of binary capture models. The MP capture scenario specifically implies
the continuous and isotropic ejection of both young and old HVSs from
the GC. {Recent observations of HVSs \citep{bro+05,hir+05,ede+05,bro+06a,bro+06b}
are consistent with a GC origin and favor a steady state temporal
distribution and an isotropic spatial distribution over a burst-like
non-spherical distribution that is expected for HVSs triggered by
the infall of a cluster \citep{lev05,haa+06,bau+06}.}

Two of the recently observed HVSs ($v\!\sim\!560$--$710\,\mathrm{km\, s^{-1}}$,
{\citealt{bro+05,bro+06a,fue+06,ede+05,bro+06b}}) were spectrally
identified as late B\,V young massive stars (masses of $3$--$5\, M_{\odot}$,
MS lifespans of $(1$--$4)\!\times\!10^{8}$ yr and number fraction
$\sim10^{-3}$ in the population), implying a total population of
$43\pm31$ such hyper-velocity stars in the Galaxy {\citep{bro+06a}}.
We model the parent binaries of these HVSs by equal mass binaries
of $M_{\mathrm{bin}}\!=\!8\,\Mo$ and $t_{\star}(4\!\Mo)\!=\!2\!\times\!10^{8}\,\mathrm{yr}$.
The ejection velocity was found in numerical simulations \citep{hil88,bro+06c}
to scale as\begin{eqnarray}
v_{\mathrm{BH}} & = & 1776\,\mathrm{km\, s^{-1}}\times\nonumber \\
 &  & \left(\frac{a}{0.1\,\mathrm{AU}}\right)^{-1/2}\left(\frac{M_{bin}}{2M_{\odot}}\right)^{1/3}\!\left(\frac{M_{BH}}{3.7\!\times\!10^{6}\,\Mo}\right)^{1/6}\,.\end{eqnarray}
 To reproduce the high HVS velocities we consider binaries with $a\!<\!1$
AU, which are tidally disrupted at $r_{t}\!<\!3.7\times10^{-4}$ pc
and eject a HVS with $v_{\mathrm{BH}}\!\gtrsim\!900\,\mathrm{km\, s^{-1}}$,
the escape velocity from the bulge \citep{haa+06}. Taking only the
GMCs into account, we predict that tens to hundreds such HVSs exist
in the Galaxy, in agreement with the deduced HVS populations, whereas
stellar relaxation predicts only 1.3 such stars (see table \ref{t:bin}).

{We note \citet{bro+06a} used the $10^{-5}\,\mathrm{yr}^{-1}$ total
rate of hyper-velocity star ejection calculated by \citet{yuq+03}
(including binaries of all stellar types, assuming stellar relaxation
only and normalized to a fiducial 10\% binary fraction)} to estimate
the number of late B\,V HVSs in a Salpeter IMF at $10$--$25$. This
theoretical prediction seems in rough agreement with the observations
(and contradicts our much lower estimate of 1.3 HVSs). However, the
rates of Yu \& Tremaine are inapplicable here and lead to a significant
over-estimate of the number of HVSs because their binary population
model is not appropriate for massive binaries in the GC. On the one
hand, the young binary population does not extend all the way to the
center, as assumed by Yu \& Tremaine for the general binary population
(Following \citealt{pau+06} who do not find any B\,V stars between
0.5--1 pc of the GC, we truncate the massive binary population inside
1.5 pc). The massive binary population in the GC is $\mathrm{few\times}10$
times smaller than implied by a simple scaling of the Yu \& Tremaine
general binary population. On the other hand, the binary fraction
of young massive binaries is 70\% rather than 10\% (\S\ref{sss:youngbin}).
We conclude that the agreement found by Brown et al. is accidental,
and that binary disruption by stellar relaxation is insufficient to
explain the number of observed HVSs, whereas MP-induced relaxation
can reproduce the observations.

\section{Summary}

\label{s:summary}

Relaxation by MPs dominates relaxation by 2-body stellar interactions
when the ratio between the 2nd moments of their respective mass functions
satisfies $\mu_{2}\!=\!\left.N_{p}\left\langle M_{p}^{2}\right\rangle \right/N_{\star}\left\langle M_{\star}^{2}\right\rangle \!\gg\!1$.
We show that Galactic MPs (stellar clusters, GMCs and smaller molecular
gas clumps that exist outside the inner few pc) dominate and accelerate
relaxation in the inner $\sim\!100$ pc of the GC. This is plausibly
the case in the centers of late-type galaxies in general. There is
also evidence for molecular gas in the centers of late-type galaxies
(e.g. \citealt{rup97,kna99}), which suggests that MPs may dominate
relaxation there as well, and lead to the relaxation of the central
regions of galactic bulges in general.

Relaxation determines the rate at which stars and binaries are deflected
to near radial (loss-cone) orbits that bring them closer to the MBH
than some critical periapse $q$, where they undergo a strong destructive
interaction with it. The size of $q$ depends on the nature of the
interaction of interest (e.g. tidal disruption, 3-body exchange).
It is much larger for binaries than for single stars due to the binary's
larger effective size.

We extend loss-cone theory to approximately treat rare encounters
with MPs, and apply it to explore the implications of MPs on the rates
of different types of close encounters. The rate reaches its maximum
when loss-cone orbits are replenished by scattering within an orbital
time (the full loss-cone regime). This is more easily achieved when
the phase-space volume of the loss-cone is small, that is, when $q$
is small. MPs thus affect only those processes with large $q$ whose
loss-cone is too large to be efficiently replenished by stellar encounters
(the empty loss-cone regime).

We show that MPs will not contribute much to the disruption of single
stars in the GC, whose loss-cone is efficiently replenished by stars
outside the central $\sim\!2$ pc (MPs may accelerate the consumption
of stars by more massive MBHs, where $q$ is significantly larger,
or the capture of stars in accretion disks). However, MPs will enhance
by factors of 10--1000 the tidal disruption rate of infalling binaries,
which result in the capture of one of the stars on a tight orbit around
the MBH, and the ejection at high velocity of the other star \citep{hil91,hil92,yuq+03}.
The enhancement of the event rates is dominated by the innermost MPs,
and so the uncertainty in the MP distribution on the smallest scales
dominates the uncertainties in the total event rate. Detailed observations
of MPs in the inner GC allow us to robustly predict their effects
in the Galaxy. We show that MP-induced disruptions of relatively rare
massive binaries can naturally explain the puzzling presence of normal-appearing
main sequence B stars in the central $0.04$ pc of the GC \citep{eis+05},
and at the same time can account for the observed HVSs well on their
way out of the Galaxy \citep{bro+05,ede+05,hir+05,bro+06a,bro+06b,bro+06c}.
Tidal disruptions of the many more faint low-mass binaries can efficiently
supply single stars on very eccentric tight orbits near the MBH. Such
an increase in the numbers of stars in tight orbit near the MBH may
increase the rates of single star processes such as tidal disruption
and heating or GW emission from tightly bound compact objects \citep{mil+05}.

Finally, MP-induced interactions also have cosmological implications
for the coalescence of binary MBHs following galactic mergers \citep{zha+02}.
We suggest that MPs can accelerate the dynamical decay of binary MBHs
by efficiently supplying stars for the slingshot mechanism, and thereby
help solve the {}``last parsec'' stalling problem. MP-driven loss-cone
refilling will operate even in the case of a spherical potential,
where other suggested mechanisms are inefficient, thus allowing MBHs
to coalesce on the dynamical timescale of the galactic merger. A detailed
treatment of this idea will be presented elsewhere (Perets \& Alexander,
in prep.).

\acknowledgements{TA is supported by ISF grant 295/02-1, Minerva grant 8563 and a New
Faculty grant by Sir H. Djangoly, CBE, of London, UK. We are grateful
to M. Tsuboi for help in interpreting the giant molecular cloud data.}

\end{document}